\documentclass[conference]{IEEEtran}

\makeatletter

\newcommand{\Rmnum}[1]{\expandafter\@slowromancap\romannumeral #1@}
\makeatother
\usepackage{url}
\usepackage{amsmath,bm}
\usepackage{amssymb}
\usepackage{extarrows}
\usepackage{algorithm}
\usepackage{graphicx} 
\usepackage{epstopdf}
\usepackage{amsthm}
\usepackage{multirow}
\usepackage{algpseudocode}
\usepackage{multicol}
\usepackage{color}
 \usepackage{cite}
\usepackage{setspace} 
\usepackage{caption}
\usepackage{subcaption}
\usepackage{subfig}
\usepackage{makecell}
\usepackage{stfloats}
\makeatletter

\theoremstyle{plain}

\algdef{SE}[DOWHILE]{Do}{doWhile}{\algorithmicdo}[1]{\algorithmicwhile\ #1}%
\begin{document}
\include{com.tex}

\title{Spectral-Efficient MIMO-OFDM: Low-Complexity Solution based on Random Multiplexing}
\author{\IEEEauthorblockN{Jie Yang$^{*}$, Wanchen Hu$^{*}$, Yi Song$^{\dagger}$, Shuangyang Li$^{\dagger}$, Burak Çakmak$^{\dagger}$, Lei  Liu$^{\ddagger}$, Xin Wang$^{*}$, and Giuseppe Caire$^{\dagger}$\\
	\IEEEauthorblockA{$^{*}$ Fudan University, Shanghai, China,  
    Emails: \{yangjie23@m., huwc23@m., xwang11@\}fudan.edu.cn\\
        $^{\dagger}$ Technische Universit\"at, Berlin, Germany.
        Emails: \{yi.song, shuangyang.li, burak.cakmak, caire\}@tu-berlin.de \\
        $^{\ddagger}$ Zhejiang University, China.
        Email: lei\_liu@zju.edu.cn\\
    }
}
}

\maketitle
\begin{abstract}
This paper presents a low-complexity precoded MIMO-OFDM system for achieving improved spectral efficiency (SE) via intentionally compressing information symbols among subcarriers. Particularly, the proposed scheme leverages the powerful random multiplexing mechanism for precoding, and adopts the linear-complexity orthogonal approximate message passing (OAMP) estimator for symbol detection, where the compatibility with the existing fifth generation (5G) architectures is fully preserved. 
We further provide the theoretical analysis based on the replica-symmetric (RS) formula. This analysis confirms the advantages of the proposed system with respect to the adopted compression ratios, where an interesting phase transition behavior is verified.
Numerical results coincide with our analysis and demonstrate significant improvements in terms of achievable rates and bit error rate (BER) compared to conventional MIMO-OFDM counterpart, making the proposed scheme a promising solution to 6G and beyond wireless networks.
\end{abstract}


%
\IEEEpeerreviewmaketitle

\section{Introduction}


As wireless communication continues to evolve toward sixth-generation (6G) systems, there is an increasing demand for improved spectral efficiency (SE) \cite{rappaportWirelessCommunicationsApplications2019,chowdhury6GWirelessCommunication2020}. Among various enabling technologies, multiple-input multiple-output orthogonal frequency division multiplexing (MIMO-OFDM) remains a fundamental transmission framework due to its ability to efficiently combat frequency-selective fading while supporting high-throughput communications \cite{tseFundamentalsWirelessCommunications}. However, under practical constraints such as limited bandwidth, finite number of antennas, and hardware complexity, achieving high SE together with low-complexity and high-performance detection remains a significant challenge.

To fully exploit the potential of MIMO systems, state-of-the-art replica-optimal signal recovery algorithms, such as approximate message passing (AMP) \cite{donoho2009message, donoho2010message}, orthogonal AMP (OAMP) \cite{maOrthogonalAMP2017b, liu2023OAMP}, vector AMP (VAMP) \cite{ranganVectorApproximateMessage2019a}, and memory AMP (MAMP) \cite{liuMemoryAMP2022}, are promising solutions. However, the theoretical analysis is generally predicated on assumptions of independent and identically distributed (i.i.d.) or right-unitarily invariant channel matrices. In practical applications, channel distributions often deviate from these assumptions, resulting in performance degradation. Recently, the interleave frequency division multiplexing (IFDM) has been proposed \cite{chiInterleaveFrequencyDivision2024}. It utilizes an inverse fast Fourier transform (IFFT) matrix cascaded with a random interleaver to construct a dense and statistically stable equivalent channel matrix, ensuring reliable signal transmission. The IFDM scheme is designed to interleave the time domain samples after IFFT, thereby resulting in an equivalent channel approximately right-unitarily invariant. A cross domain MAMP detector has been proposed for the IFDM scheme to reduce the detection complexity while approaching the performance of the OAMP. {Moreover, the interleaving can be extended to the time, frequency, and spatial domains \cite{yan2026capacity}, providing a coding framework for a sparse regression code achieving the regional capacity. However, in large-scale systems, high-dimensional cross-domain transform and detection may face high complexity challenges in hardware implementation \cite{liuInterleavedBlockSparseTransform2025}.}

In addition to advanced detection algorithms, complexity-scalable MIMO transceiver design has been extensively investigated as an effective alternative to improve SE in MIMO systems. By jointly optimizing precoding, modulation, and detection strategies, significant performance gains can be achieved under practical constraints. In particular, techniques such as iterative transceiver optimization and bit interleaved coded modulation (BICM)-based designs can approach the achievable rate limits by better exploiting the spatial degrees of freedom. Our recent works reveal that the optimal precoding is significantly dependent on the signal-to-noise ratio (SNR) \cite{yang2025achievable,yang2026complexity}.
Nevertheless, most of the existing MIMO-OFDM implementations are based on Nyquist transmissions, which 
limits the number of symbols to be transmitted given the available resources, e.g., the antenna configuration and the number of subcarriers. Consequently, the SE of practical MIMO-OFDM systems is typically suboptimal due to the adoption of non-Gaussian constellations. 
This calls for advanced shaping schemes for MIMO-OFDM with practical non-Gaussian constellations to further improve the SE.

In this paper, we introduce the spectral-efficient MIMO-OFDM (SE-MIMO-OFDM) transmission, which is a scheme inspired by the recently proposed random multiplexing \cite{liuRandomMultiplexing2026}.
The key idea is to introduce a randomized linear transformation/compression in the frequency domain before the transmission of conventional MIMO-OFDM, which effectively reshapes the effective frequency domain channel after conventional MIMO-OFDM receiver processing into a statistically isotropic form, thereby eliminating dominant directions in the signal space. 
Particularly, this enables a rigorous analytical simplification, whereby the high-dimensional detection problem decouples into several equivalent scalar Gaussian channels characterized by the identical effective SNR. Building on this decoupling principle, the system performance can be precisely characterized via a set of self-consistent fixed-point equations derived using the replica method, which aligns with the state evolution of OAMP/VAMP algorithms. As a result, near-optimal Bayesian inference can be achieved with tractable computational complexity.

{From a system perspective, the proposed SE-MIMO-OFDM is designed as a modular add-on to conventional MIMO-OFDM architectures}. Specifically, frequency domain data symbols are mapped onto available subcarriers through a linear compression layer prior to standard MIMO-OFDM transmission. Then, an OAMP estimator is adopted to further refine the estimates of frequency domain outputs after the receiver processing of conventional MIMO-OFDM.
Importantly, the proposed scheme preserves the original MIMO-OFDM signaling structure as well as existing MIMO processing techniques, such as SVD-based spatial precoding, enabling seamless integration into current 5G and emerging 6G systems with minimal implementation overhead. Numerical results demonstrate that the proposed scheme significantly outperforms conventional MIMO-OFDM systems, achieving notable spectral efficiency gains while maintaining low computational complexity. These advantages make the proposed scheme a promising candidate for next-generation wireless systems.

\section{System Model} \label{SEC2}

\begin{figure*}
    \centering
    \includegraphics[width=0.8\linewidth]{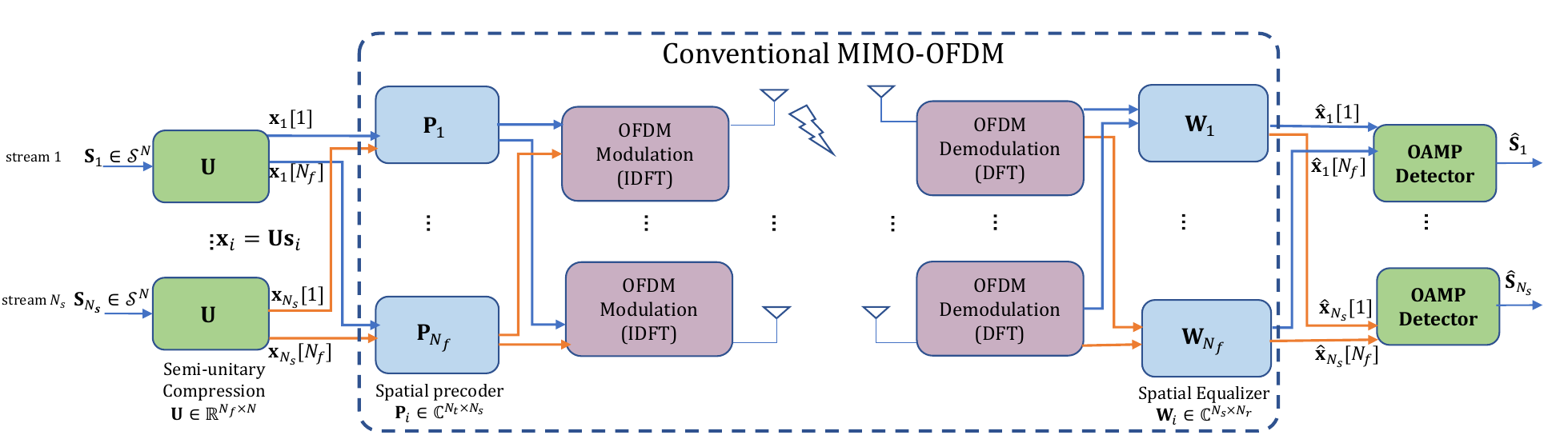}
    \caption{The proposed SE-MIMO-OFDM transmission system model. }
    \label{fig:systemmodel}
\end{figure*}
\subsection{Conventional MIMO-OFDM System}
\label{sec:conv_MIMO_OFDM}
Consider a standard MIMO-OFDM system with $N_f$ subcarriers. Let $\mathbf{H}_k \in \mathbb{C}^{N_r \times N_t}$ denote the spatial channel matrix of the $k$-th subcarrier, with $N_t$ and $N_r$ representing the number of transmit and receive antennas, respectively. Here, the spatial and frequency dimensions are decoupled to facilitate per-subcarrier processing.  We assume that the transmitted data stream is mapped onto $N_s \le \min\{N_r, N_t\}$ layers, and the corresponding frequency-domain symbol matrix is denoted by $\mathbf{X} \in \mathbb{C}^{N_f \times N_s}$. To maximize the SE, spatial precoding based on singular value decomposition (SVD) is typically employed. For the $k$-th subcarrier, the channel is decomposed as $\mathbf{H}_k = \mathbf{\Gamma}_k \boldsymbol{\Lambda}_k \mathbf{V}_k^H$.

Following the standard MIMO-OFDM workflow, i.e., the modulation (including normalized IFFT and cyclic prefix (CP) insertion) and demodulation (including CP removal and FFT), the received signal at the $k$-th subcarrier is expressed as:\begin{equation}\label{eq:precoding}\Bar{\mathbf{y}}_k = \mathbf{H}_k \mathbf{P}_k \Bar{\mathbf{x}}_k + \Bar{\mathbf{z}}_k, \quad k=1, 2, \dots, N_f,\end{equation}
where $\Bar{\mathbf{x}}_k \in \mathbb{C}^{N_s \times 1}$ is the transmitted symbol vector across $N_s$ spatial layers corresponding to the $k$-th subcarrier (the $k$-th column of $\mathbf{X}^T$), $\mathbf{P}_k \in \mathbb{C}^{N_t \times N_s}$ is the precoding matrix consisting of the first $N_s$ columns of $\mathbf{V}_k$, and $\Bar{\mathbf{z}}_k \sim \mathcal{CN}(\mathbf{0}, N_0\mathbf{I}_{N_s})$ is the additive white Gaussian noise (AWGN). 

By applying the matching filter $ \mathbf{W}_k = \mathbf{\Gamma}_k^H$ at the receiver, one transforms the MIMO channel into $N_s$ parallel spatial layers. The post-equalization signal for the $i$-th layer on the $k$-th subcarrier can be written as:
\begin{equation}\label{eq:mimoy}{y}_{k,i} = \lambda_{k,i} x_{k,i} + {z}_{k,i}, \quad i=1, \dots, N_s,
\end{equation}
where $\lambda_{k,i}$ is the $i$-th singular value of $\mathbf{H}_k$, and $x_{k,i}$ is the symbol transmitted on the $k$-th subcarrier and the $i$-th layer. ${z}_{k,i}$ denotes the effective noise. Consequently, for the energy-normalized symbol $x_{k,i}$, the effective SNR for each resource element is defined as $1/\sigma^2_{k,i} = \lambda^2_{k,i}/N_0$. Notably, this effective SNR fluctuates across different subcarriers and spatial layers due to the frequency-selective fading and the inherent power variations of the spatial eigen-modes.

\subsection{Proposed SE-MIMO-OFDM Model}

In this work, we propose a novel transmission scheme that enhances SE by introducing two additional modules into the standard
MIMO-OFDM architecture, as illustrated in Fig.~\ref{fig:systemmodel}. The
key distinction from the conventional system lies in the construction of the
frequency-domain symbol matrix~$\mathbf{X}$.

Specifically, the transmitter sends one data stream, which goes through layer mapping and obtains an information symbol matrix $\mathbf{S} \in \mathbb{C}^{N \times N_s}$, whose $(n, i)$-th entry $s_{n,i}$ denotes the $n$-th information symbol associated to the $i$-th layer. Furthermore, $N$ is the number of information symbols per layer. Let $\mathbf{s}_i = [s_{1,i}, s_{2,i},\cdots,s_{N,i}]^T$ be the data vector in the $i$-th layer, where each entry is drawn independently from a complex constellation $\mathcal{S}$ (e.g., QPSK or $M$-QAM) with unit average power, i.e., $\mathbb{E}[\lvert s_{n,i} \rvert^2] = 1$. To perform frequency-domain compression, i.e., mapping $N$ information symbols onto a reduced set of $N_f$ subcarriers ($N_f \leq N$), we adopt the idea of random multiplexing~\cite{liuRandomMultiplexing2026} by employing a
semi-unitary matrix $\mathbf{U} \in \mathbb{C}^{N_f \times N}$ satisfying
$\mathbf{U}\mathbf{U}^H = \mathbf{I}_{N_f}$. Crucially, $\mathbf{U}$ is
drawn from the Haar measure on the set of semi-unitary matrices and is
channel-agnostic, meaning it requires no knowledge of the channel state
information. Note that $\mathbf{U}^H\mathbf{U} \neq \mathbf{I}_{N}$ due to
the dimensionality reduction, when the compression ratio $\beta = N_f/N$ is smaller than $1$. The resultant frequency-domain symbol matrix
$\mathbf{X} \in \mathbb{C}^{N_f \times N_s}$ is then given by
\begin{equation}\label{eq:2}
  \mathbf{X} = \mathbf{U}\mathbf{S}.
\end{equation}

Consistently with \eqref{eq:mimoy}, the model for the $i$-th layer over the $N_f$ subcarriers is 
\begin{equation}\label{eq:equaly}
    {\mathbf{y}}_i  = \mathbf{\Lambda}_i \mathbf{x}_i + {\mathbf{z}}_i,
\end{equation}
where ${\mathbf{z}}_i \sim \mathcal{CN}(\mathbf{0},N_0\mathbf{I}_{N_f})$, and

\begin{equation}\label{eq:estx}
    \mathbf{\Lambda}_i = \mathrm{diag}\left(\left[\lambda_{1,i},\lambda_{2,i},\cdots,\lambda_{N_f,i}\right]^T\right).  
\end{equation}
Dropping the layer index~$i$ for brevity, \eqref{eq:equaly} is re-expressed as
\begin{equation}
  \mathbf{y}\triangleq \mathbf{\Lambda}\mathbf{U}\,\mathbf{s} + \mathbf{z} = \mathbf{A}\,\mathbf{s} + \mathbf{z}, \label{eq:finalmodel}
\end{equation}
where $\mathbf{A} \triangleq \mathbf{\Lambda}\mathbf{U}$ denotes the effective channel matrix that jointly captures the frequency-selective channel and the compression operation, and $\mathbf{z} \sim \mathcal{CN}(\mathbf{0}, N_0\mathbf{I}_{N_f})$ represents the AWGN. The recovery of the information symbol vector $\mathbf{s}$ is then carried out by the OAMP detector based on the unified linear model in \eqref{eq:finalmodel}. Consequently, the SE of the proposed scheme is given as
\begin{equation}\label{eq:se}
    {\rm SE} = \frac{N_f\cdot \mathcal{I}/\beta}{T B} {\rm bits/s/Hz/Stream},
\end{equation}
where $B = N_f \cdot \Delta f$ is the bandwidth with the frequency interval $\Delta f$, and $T = 1/\Delta f$. $\mathcal{I}$ is the mutual information (MI) per symbol per stream, which will be investigated later.



The proposed SE-MIMO-OFDM framework offers two key advantages. First, since
$\mathbf{U}$ is channel-agnostic, the scheme is fully compatible with existing physical-layer architectures; the compression and subsequent OAMP detection can be incorporated as add-on modules, enabling a seamless evolution toward next-generation high-capacity networks. Second, the framework provides a flexible trade-off between throughput and reliability through the compression ratio $\beta$. {It is worth noting that, due to the ordered eigen-subchannels, the effective SNRs across different layers are inherently imbalanced, resulting in heterogeneous SEs. To better exploit the available degrees of freedom, adaptive selection of $\beta$ together with modulation and coding schemes (MCSs) per-layer is essential for approaching the channel capacity region with Gaussian signals. This observation is consistent with the classical water-filling principle, where different eigen-modes support different achievable rates, which will be investigated for the coded system in our future work.}

\section{Replica Prediction of Mutual Information}
\begin{figure*}	
  \centering
	\begin{subfigure}[htbp]{2.35in}
		
		\includegraphics[width=2.35in]{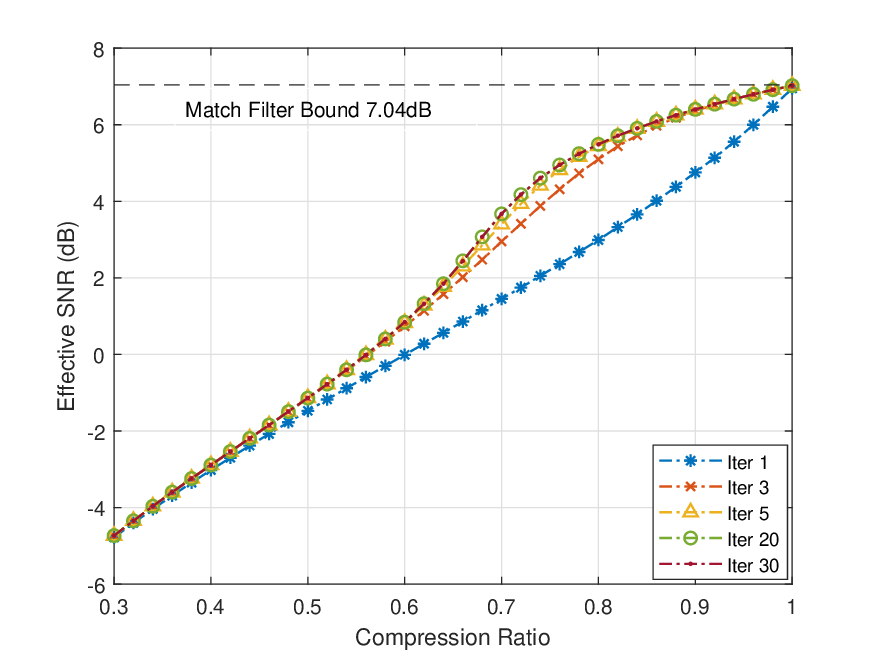}
		\caption{Input SNR $-5$ dB.}\label{fig.map}		
	\end{subfigure}
	\begin{subfigure}[htbp]{2.35in}
		
		\includegraphics[width=2.35in]{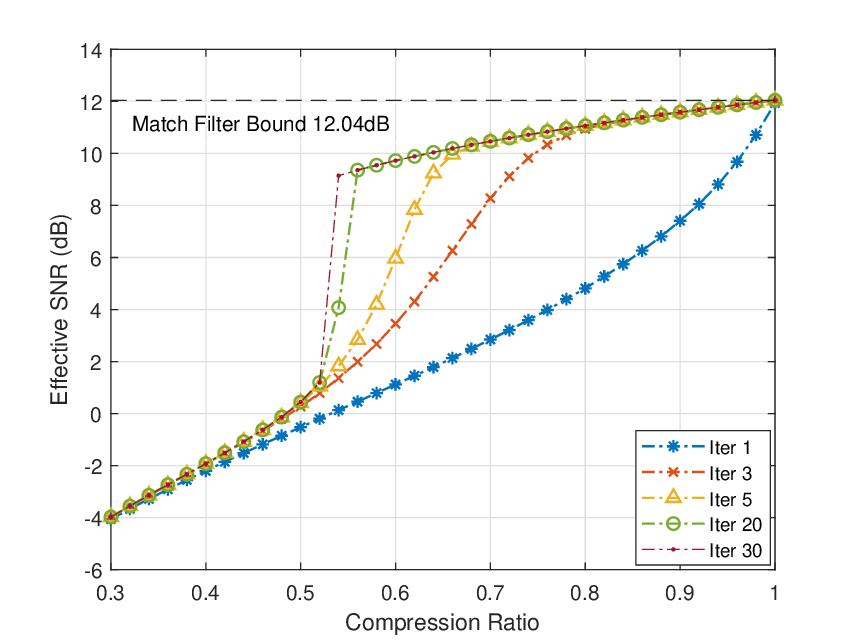}
		\caption{Input SNR $0$ dB.}\label{fig.cond}
	\end{subfigure}
    	\begin{subfigure}[htbp]{2.35in}
		
		\includegraphics[width=2.35in]{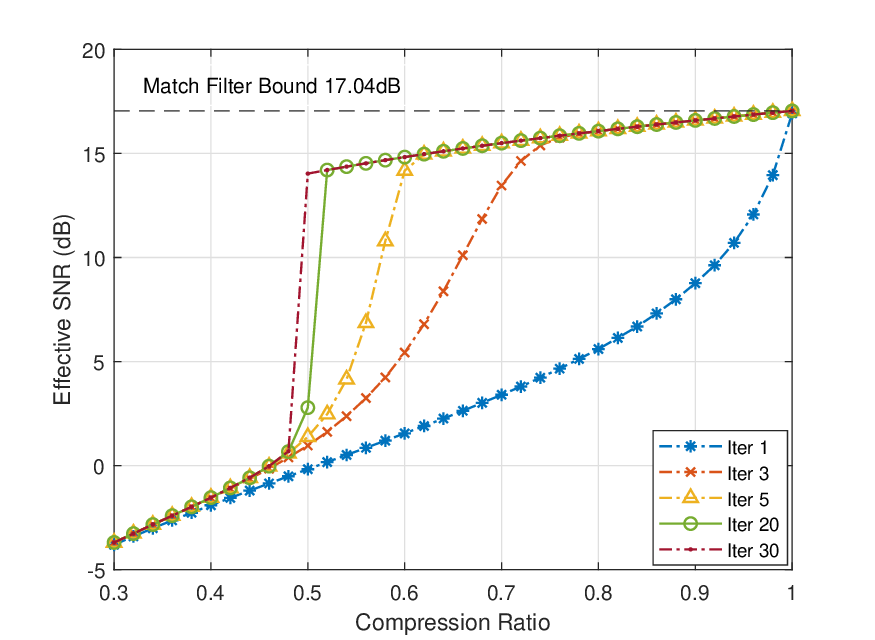}
		\caption{Input SNR $5$ dB.}\label{fig.scatter}		
	\end{subfigure}
	\caption{Effective SNR $\rho$ vs. compressed ratio $\beta$ derived for uncoded QPSK, where different input SNR are considered.}\label{fig.1}
\end{figure*}
To evaluate the theoretical limit of the proposed SE-MIMO-OFDM system, we analyze its asymptotic MI in the large system limit, where $N, N_f \to \infty$. {Specifically, we are interested in computing 
the asymptotic normalized input-output MI:
\begin{equation}
\begin{aligned}
    \mathcal{I} &= \lim_{N \to \infty}\frac{1}{N} \mathcal{I}(\mathbf{s};\mathbf{y}|{\bf A}) =\lim_{N \to \infty}\frac{1}{N}\mathbb{E}\left[\log_2\frac{p(\mathbf{y}|\mathbf{s})}{p(\mathbf{y})}|{\bf A}\right].
\end{aligned}
\end{equation}


For convenience, we introduce for $x\in(0,\infty)$
\begin{subequations}
 \begin{align}
\mathcal I_1(x)&\triangleq \lim_{N\to \infty} \frac{1}{N}\mathcal I\left(\mathbf s;\sqrt{x}\mathbf s+\mathbf w\right) \\
\mathcal I_2(x)&\triangleq
\lim_{N\to \infty}\frac{1}{N}\log_2\det(x\mathbf I_{N}+{\rm snr}\mathbf A^H \mathbf A)\;.
\end{align}   
\end{subequations}
Here, ${\rm snr} = 1/N_0$ and $\mathbf{w}\sim \mathcal{CN}(\mathbf{0},\mathbf{I}_N)$.
Then, by means of the replica-symmetric (RS) analysis \cite{tanakaStatisticalmechanicsApproachLargesystem2002,tulinoSupportRecoverySparsely2013}, the MI can be expressed as \cite{RFTN}
\begin{align}
 \mathcal I\overset{{\rm RS}}{=}\inf_{\Theta}~\big[\mathcal I_1(\rho)+\mathcal I_2(\nu)-\log_2(\rho+\nu) \big],\label{replica_MI}
\end{align}
where the infimum is defined over the set
\begin{align}
\Theta\triangleq \left\{\rho,\nu\geq 0:\mathcal I'_1(\rho)=\mathcal I'_2(\nu)=(\rho+\nu)^{-1}
\right\}\;. \label{theta}
\end{align}
Here, $\mathcal I_{1}'$ and $\mathcal I_{2}'$ denote the derivatives of $\mathcal I_{1}$ and $\mathcal I_{2}$, respectively and they are given by
\begin{subequations}
\begin{align}
{\mathcal  I}'_1(x)&\triangleq\lim_{N\to\infty}\frac{1}{N}{\mathbb E}\left[\Vert \mathbf s-{\mathbb E}\left[\mathbf s| \sqrt{x}\mathbf s+\mathbf w\right]\Vert^2\right],\\
{\mathcal I}'_2(x)&\triangleq  \lim_{N\to \infty}\frac{1}{N}{\rm tr}((x\mathbf I_N+{\rm snr}\mathbf A^H \mathbf A)^{-1})\;.
\end{align}
\end{subequations}

Let $(\rho^\star,\nu^\star)$ stand for the global minimum solution in \eqref{replica_MI}. Then, $\rho^\star$ can be interpreted as \emph{effective SNR} for the equivalent (w.r.t. the asymptotic analysis) input-output decoupled system $(\mathbf s;\sqrt{\rho^\star}\mathbf s+\mathbf z)$.  For example, the asymptotic normalized mean-square error predicted by the RS ansatz is given by
\begin{equation}
   \lim_{N\to \infty}\frac{1}{N}\mathbb E[\Vert{\mathbf s-\mathbb E[\mathbf s\vert \mathbf y,\mathbf A]}\Vert^2\vert \mathbf A] \overset{\rm RS}{=} \mathcal I_1'(\rho^\star).  
\end{equation}

We now define an iterative process for solving the fixed-point equations given in \eqref{theta}, i.e., 
\begin{equation}
\mathcal I'_1(\rho)=\mathcal I'_2(\nu)=(\rho+\nu)^{-1}\;. \label{fixed_point_equation}
\end{equation}
Specifically, we begin with the initialization $\nu^{(1)}=\frac 1{\mathcal I_1'(0)} = 1$ and we then iteratively update for $t=1,2,...$
\begin{subequations} \label{fixed_point_solution}
\begin{align}
\rho^{(t)} &=\frac{1}{\mathcal I_2'(\nu^{(t)})}-\nu^{(t)}\label{solution2}, \\
\nu^{(t+1)} &= \frac{1}{\mathcal I_1'(\rho^{(t)})}- \rho^{(t)}.
\end{align}
\end{subequations}
As $t \to \infty$, it can be shown that $\left(\nu^{\left(t\right)},\rho^{\left(t\right)}\right)$ converges to the fixed-point solution of~\eqref{fixed_point_equation}. In particular, in the so-called algorithmic phase (see \cite{barbier2019optimal}), where the RS fixed-point equation admits a unique solution, the above iterative scheme can be used to compute the MI ~\eqref{replica_MI}. In fact, this iteration coincides with the so-called \textit{state evolution} of OAMP/VAMP decoding algorithm.}

As an illustration, Fig. 2 shows the effective SNR $\rho$ as a function of the compression ratio $\beta$, where $\beta\leq 1$. Numerical results are obtained by averaging on $N_f = 1024$ subchannels. Each subchannel has $N_r =1$ and $N_t = 16$. 
A clear phase transition behavior is observed with respect to $\beta$, i.e.,  when $\beta$ falls below a certain threshold, the effective SNR $\rho$ degrades sharply, leading to a significant performance loss in signal recovery. Moreover, this phase transition can be alleviated by increasing the input SNR, indicating that higher input SNR enables reliable recovery at lower compression ratios. These observations highlight a fundamental trade-off between the input SNR and the compression ratio. In particular, properly balancing these two factors is critical to pushing the phase transition threshold toward lower $\beta$, which will be investigated in our journal paper.

\section{OAMP Detector}
\subsection{OAMP Algorithm}
The proposed SE-MIMO-OFDM leverages the OAMP detector to recover the high-dimensional signal $\mathbf{s}$ from the compressed observations $\mathbf{y}$. OAMP iteratively refines the estimate through two decoupled stages: a linear estimation (LE) step and a non-Linear estimation (NLE) step, ensuring the error remains Gaussian-distributed at each iteration. This is particularly suitable for our scheme, because the specific structure of the compression matrix $\mathbf{AA}^H = \mathbf{\Lambda}^{2}$ allows for an effective implementation with significantly reduced computational complexity. Starting from $\mathbf{s}^{(0)} = \mathbf{0}$ and $\eta^{(0)} = 1$, the algorithm proceeds as follows:

1) \textbf{LE Step: Linear decoupling and Onsager correction}. The LE step aims to decouple the mixed signals and provide an extrinsic estimate for the NLE.
\begin{subequations}
\begin{align}{\mathbf{\Phi}}^{(t)} &= 
\text{diag}\left(\left[\frac{1}{\eta^{(t)}\lambda_{1}^2 + N_0}, \cdots, \frac{1}{\eta^{(t)}\lambda_{N_f}^2 + N_0}\right]\right), \label{subeq:phi}\\
{\chi^{(t)}} &= \frac{1}{N} \sum_{k=1}^{N_f} \frac{1}{\eta^{(t)} + \sigma_k^2}, \label{subeq:nu}\\
\rho^{(t)} &= (1/{\chi^{(t)}}-\eta^{(t)})^{-1}, \label{subeq:rho}\\
\mathbf{r}^{(t)} &= \mathbf{s}^{(t)} + \frac{1}{{\chi^{(t)}}}\mathbf{A}^H\mathbf{\Phi}^{(t)}(\mathbf{y-As}^{(t)}). \label{subeq:r}\end{align}\end{subequations}
Here, \eqref{subeq:phi} computes the LMMSE filtering matrix based on the current prior variance $\eta^{(t)}$, which suppresses both the additive noise and the inter-symbol interference; \eqref{subeq:nu} calculates the average posterior error variance of the LMMSE estimate, reflecting the noise reduction capability of the linear stage. By removing the correlation between the current estimate and the prior input, \eqref{subeq:rho} characterizes the extrinsic noise of the decoupled observation. Finally, \eqref{subeq:r} generates the de-biased observation $\mathbf{r}^{(t)}$, which can be modeled as a signal corrupted by AWGN with variance $1/\rho^{(t)}$. { As a matter of fact, $\rho^{(t)}$ here is algorithmic computation (for large $N$) of the asymptotic quantity $\rho^{(t)}$ in \eqref{fixed_point_solution} which is used for solving  the replica fixed-point equations.\color{black}}

2) \textbf{NLE Step: Constellation-aware denoising}. The NLE step utilizes the discrete nature of the constellation $\mathcal{S}$ to refine the estimate.
\begin{subequations}
\begin{align}
\tau^{(t+1)} &= \frac{1}{N} \mathbb{E} \left[ \|\mathbf{s} - \mathbb{E}[\mathbf{s} | \sqrt{\rho^{(t)}}\mathbf{s} + \mathbf{w}]\|^2 \right],\label{subeq:chi} \\
\hat{\mathbf{s}}^{(t+1)} &= \mathbb{E}[\mathbf{s}|\mathbf{r}^{(t)} = \mathbf{s} +\frac{1}{\sqrt{\rho^{(t)}}} \mathbf{w}],\label{subeq:s_hat}\\
\mathbf{s}^{(t+1)} &= \frac{\frac{\hat{\mathbf{s}}^{(t+1)}}{\rho^{(t)}}- \tau^{(t+1)}\mathbf{r}^{(t)}}{1/\rho^{(t)}-\tau^{(t+1)}},\label{subeq:s_next}\\
\eta ^{(t+1)}&=(1/\tau^{(t+1)}-\rho^{(t)})^{-1} \label{subeq:eta_next}.
\end{align}
\end{subequations}
Here, \eqref{subeq:chi} computes the MMSE of the scalar Gaussian denoising problem, which serves as the new posterior error variance; \eqref{subeq:s_hat} applies the posterior mean estimator based on the constellation $\mathcal{S}$, effectively projecting the noisy observations back to the valid signal space. To prevent the propagation of self-feedback, \eqref{subeq:s_next} extracts the extrinsic signal $\mathbf{s}^{(t+1)}$ as the prior for the next LE step. Finally, \eqref{subeq:eta_next} updates the prior variance $\eta^{(t+1)}$ for the next iteration.

\subsection{Complexity Analysis}

The computational complexity of the OAMP-based detector is primarily governed by the LE step. In this section, we demonstrate how the inherent structure of the proposed SE-MIMO-OFDM framework significantly alleviates the processing burden. 

For a general linear model $\mathbf{y} = \mathbf{A}\mathbf{s} + \mathbf{w}$, the LE step conventionally requires the computation of a LMMSE matrix $
\mathbf{W} = \mathbf{A}^H (\mathbf{I} + \eta \mathbf{A}\mathbf{A}^H)^{-1} $.
In a standard implementation, the inversion of the $N_f \times N_f$ matrix incurs a cubic complexity of $\mathcal{O}(N_f^3)$, while matrix-vector multiplications require $\mathcal{O}(N_f N)$ operations per iteration. Therefore, the overall complexity of OAMP scales cubically with the system dimension in the general case, which may become prohibitive for large-scale systems. In contrast, in the proposed SE-MIMO-OFDM scheme, the semi-unitary property of the compression matrix ($\mathbf{AA}^H = \mathbf{\Lambda}^{2}$) and the diagonalized noise covariance matrix $\boldsymbol{\Sigma}$ enable a closed-form simplification. As shown in \eqref{subeq:phi}, the matrix inversion is reduced to a diagonal-form operation. Consequently, the inversion complexity collapses from $\mathcal{O}(N_f^3)$ to $\mathcal{O}(N_f)$. The remaining dominant operation is the linear transformation $\mathbf{A}^H \mathbf{\Phi}^{(t)}$, which can be efficiently implemented, resulting in a total LE complexity that scales linearly with the system dimensions.
{The NLE} of OAMP operates independently symbol by symbol and involves evaluating posterior probabilities over the constellation set. Its complexity scales linearly with the number of symbols, i.e., $\mathcal{O}(N|\mathcal{S}|)$, where $|\mathcal{S}|$ denotes the constellation size.

In summary, by leveraging the semi-unitary structure of $\mathbf{U}$ and the inherent independent layers in MIMO-OFDM systems, the proposed SE-MIMO-OFDM achieves a per-iteration complexity that scales linearly with the system dimension. This represents a substantial reduction compared to conventional OAMP implementations, making the proposed scheme well-suited for practical wireless systems.

\section{Numerical Results}
\begin{figure}
    \centering
    \includegraphics[width=1.0\linewidth]{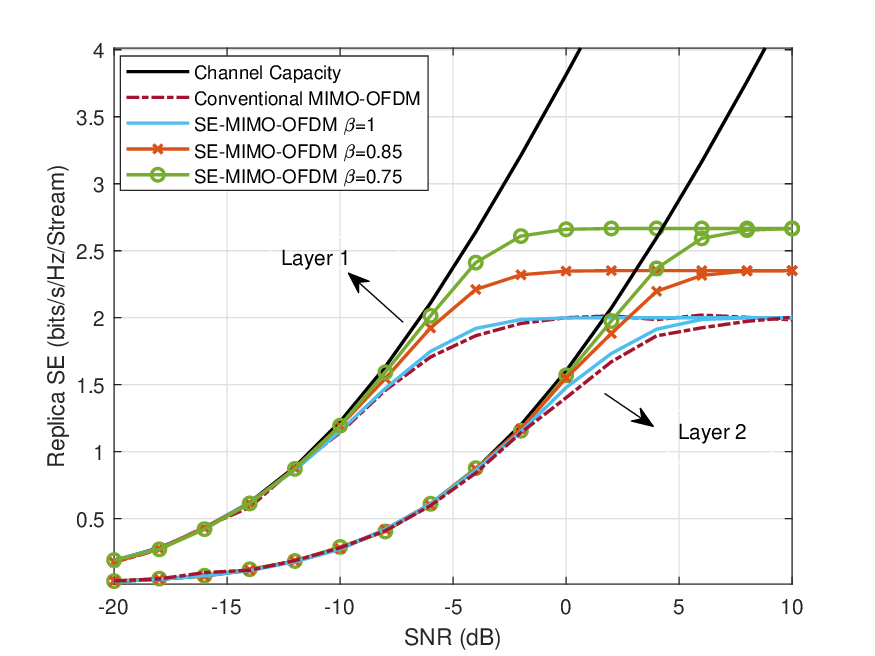}
    \caption{The replica MI of $2\times16$ MIMO with $N_s=2$ layers QPSK transmission. }
    \label{fig:MI}
\end{figure}
\begin{figure}
    \centering
    \includegraphics[width=1.0\linewidth]{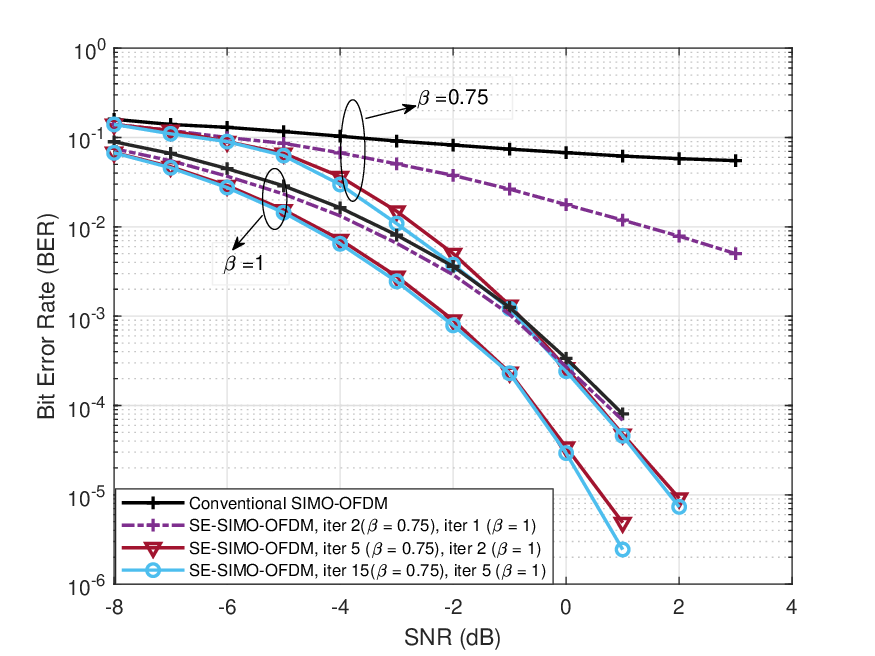}
    \caption{$1\times16$ SIMO with $N_s=1$ layer QPSK transmission.}
    \label{fig:betasimo}
\end{figure}
\begin{figure}
    \centering
    \includegraphics[width=1.0\linewidth]{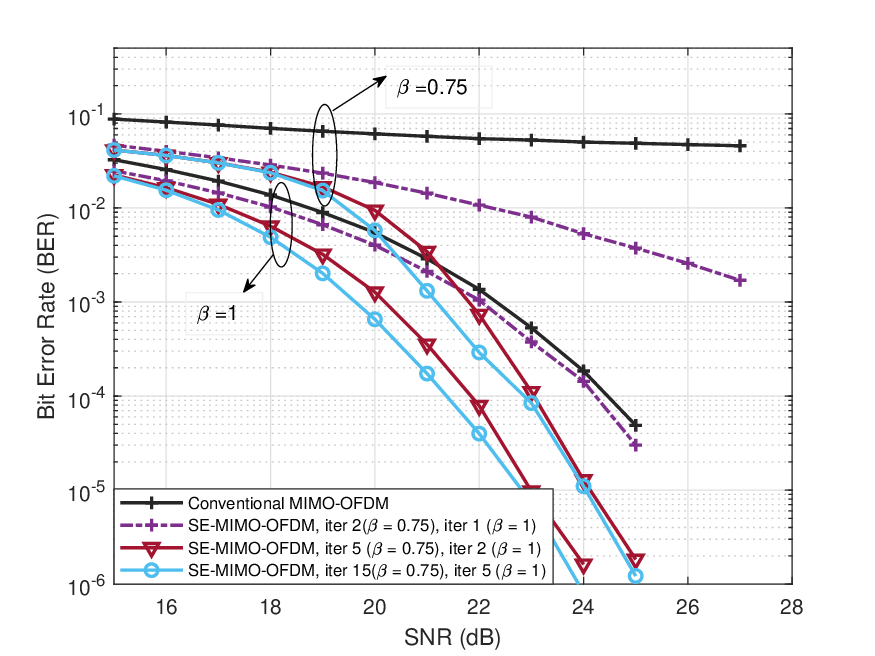}
    \caption{$4\times16$ MIMO with $N_s = 4$ layers QPSK transmission.}
    \label{fig:betamimo}
\end{figure}

In this section, we provide the numerical results in different SE-MIMO-OFDM systems with QPSK transmission, respectively. The conventional MIMO-OFDM scheme is adopted as the baseline. Let $\left\{\mathbf{H}_k\right\}_{k=1}^{N_f}$ (cf. (\ref{eq:precoding})) be the frequency response of a MIMO-OFDM channel with $L$ paths and $N_f$ subcarriers. Here, a widely used Saleh-Valenzuela mmWave channel model \cite{huParametricMIMOOFDMChannel2025a} is considered. We set the number of multipaths $L = 10$ and the other channel parameters are generated according to the 3GPP 38.901 specification \cite{38901}, where we set $N_f = 1024$. The SNR is defined as $10\log_{10}(1/N_0)$.

{Fig.~\ref{fig:MI} illustrates the SE of two separate layers in a $2\times 16$ MIMO system with QPSK signaling according to \eqref{eq:se}. Conventional OFDM saturates at $\log_2|\mathcal{S}|$ bits per symbol, leading to a pronounced \emph{shaping loss} at high SNR. Here, $\mathcal{S}$ is the QPSK for illustration. In contrast, the proposed scheme with $\beta<1$ achieves a clear MI gain in the low-to-moderate SNR regime. In particular, for $\beta=0.75$, the MI closely approaches the Gaussian-input capacity before saturation at $\log_2|\mathcal{S}|/\beta$. In addition, due to the ordered eigenvalues of the MIMO channel, the two spatial layers exhibit distinct SNR thresholds, indicating that a per-layer adaptive modulation and coding scheme (MCS) is preferable compared to a uniform design in the practical coded system left for future consideration.}

{The above performance gain originates from the structural effect of the semi-unitary transformation. Specifically, the proposed scheme maps $N$ symbols onto $N_f < N$ subcarriers via a Haar-distributed matrix $\mathbf{U}$, which introduces a higher-dimensional coupling across subcarriers. This coupling enables one to effectively re-distribute the excess capacity of strong subcarriers to weaker ones. As a result, the shaping loss is reduced in the moderate SNR regime.}

{Fig.~\ref{fig:betasimo} evaluates the bit error rate (BER) performance of the proposed scheme under compression ratios $\beta = \{0.75,  1\}$ in the $1 \times 16$ SIMO system with $N_s = 1$ data stream, and Fig.~\ref{fig:betamimo} presents the corresponding results for the $4 \times 16$ MIMO system with $N_s = 4$ layers. In both figures, the conventional MIMO-OFDM baseline employs an LMMSE detector for signal recovery.
A notable observation is that the proposed SE-MIMO-OFDM with OAMP significantly outperforms the conventional baseline in the non-compressed case ($\beta = 1$). The fundamental reason is that the optimality of the conventional MIMO-OFDM scheme relies on the Gaussian inputs. For practical finite-alphabet constellations such as QPSK, the detector fails to exploit the discrete structure of the transmitted symbols, exhibiting limited performance.

{The proposed scheme closes this gap through the OAMP detector, whose Bayes-optimality is enabled by the Haar-distributed compression matrix $\mathbf{U}$. Specifically, the Haar structure ensures that the effective channel matrix $\mathbf{A} = \boldsymbol{\Lambda}\mathbf{U}$ is right-rotationally invariant, which allows the OAMP to decouple the high-dimensional estimation problem into a sequence of equivalent scalar Gaussian channels. At each iteration, the LE step performs an LMMSE operation in the transformed domain with low complexity, while the NLE step applies the Bayes-optimal MMSE denoiser that exploits the prior knowledge of the constellation $\mathcal{S}$. By iterating between these two steps, OAMP has been proven to achieve the Bayes-optimal performance predicted by the replica method. In the compressed case $\beta = 0.75$, the proposed scheme also exhibits improved BER performance compared to the conventional MIMO-OFDM system, while transmitting more information bits. It is also observed that, in the compressed case, the information recovery in the conventional MIMO-OFDM system is difficult due to the low-rank measurement. These results show that the proposed SE-MIMO-OFDM scheme is a promising candidate technique in next generation communication.}

\section{Conclusion}
In this paper, we presented a high-compatibility transmission framework designed to improve the SE limits of conventional MIMO-OFDM systems without overhauling the existing physical layer architecture. Our approach operates as a seamless add-on module that embeds supplementary data into the available subcarriers through a linear transformation in the frequency domain. The core contribution of this work is the demonstration that substantial SE gains can be achieved while maintaining full transparency to standard 5G procedures or future 6G systems, including SVD-based precoding and per-subcarrier equalization. By exploiting the structural properties of our proposed precoding, we showed that reliable signal recovery can be implemented with a low computational complexity using OAMP. Our asymptotic analysis and numerical results further confirm that this framework provides a practical and cost-effective pathway for enhancing the capacity of future wireless networks, offering a robust balance between theoretical performance gains and engineering feasibility.
\bibliographystyle{ieeetr}
\bibliography{all}

@article{barbier2019optimal,
  title={Optimal errors and phase transitions in high-dimensional generalized linear models},
  author={Barbier, Jean and Krzakala, Florent and Macris, Nicolas and Miolane, L{\'e}o and Zdeborov{\'a}, Lenka},
  journal={Proc. Natl. Acad. Sci. USA},
  volume={116},
  number={12},
  pages={5451--5460},
  year={2019},
  publisher={National Academy of Sciences}
}

@article{chowdhury6GWirelessCommunication2020,
  title = {{{6G Wireless communication systems}}: {{Applications}}, {{requirements}}, {{technologies}}, {{challenges}}, and {{research directions}}},
  author = {Chowdhury, Mostafa Zaman and Shahjalal, {\relax Md}. and Ahmed, Shakil and Jang, Yeong Min},
  year = 2020,
  journal = {IEEE Open Journal of the Communications Society},
  volume = {1},
  pages = {957--975},
  issn = {2644-125X},
  doi = {10.1109/OJCOMS.2020.3010270},
  urldate = {2026-04-07}
}

@inproceedings{donoho2010message,
  title={{Message passing algorithms for compressed sensing: I. Motivation and construction}},
  author={Donoho, David L and Maleki, Arian and Montanari, Andrea},
  booktitle={Proceedings of IEEE Information Theory Workshop (ITW 2010, Cairo)},
  pages={1--5},
  year={2010},
  organization={IEEE}
}

@article{donoho2009message,
  title={{Message-passing algorithms for compressed sensing}},
  author={Donoho, David L and Maleki, Arian and Montanari, Andrea},
  journal={Proceedings of the National Academy of Sciences},
  volume={106},
  number={45},
  pages={18914--18919},
  year={2009},
  publisher={National Academy of Sciences}
}

@ARTICLE{liu2023OAMP,
  author={Liu, Lei and Cheng, Yiyao and Liang, Shansuo and Manton, Jonathan H. and Ping, Li},
  journal={IEEE Trans. Commun.}, 
  title={On {OAMP}: {Impact} of the Orthogonal Principle}, 
  year={2023},
  volume={71},
  number={5},
  pages={2992-3007}}

@article{rappaportWirelessCommunicationsApplications2019,
  title = {Wireless {{communications}} and {{applications Above}} 100 {{GHz}}: {{Opportunities}} and {{challenges}} for {{6G}} and {{beyond}}},
  author = {Rappaport, Theodore S. and Xing, Yunchou and Kanhere, Ojas and Ju, Shihao and Madanayake, Arjuna and Mandal, Soumyajit and Alkhateeb, Ahmed and Trichopoulos, Georgios C.},
  year = 2019,
  journal = {IEEE Access},
  volume = {7},
  pages = {78729--78757},
  issn = {2169-3536},
  doi = {10.1109/ACCESS.2019.2921522},
  urldate = {2026-04-07}
}

@book{tseFundamentalsWirelessCommunications,
  title={{Fundamentals of Wireless Communication}},
  author={Tse, David and Viswanath, Pramod},
  year={2005},
  publisher={Cambridge university press}
}

@article{chiInterleaveFrequencyDivision2024,
  title = {Interleave {{frequency division multiplexing}}},
  author = {Chi, Yuhao and Liu, Lei and Ge, Yao and Chen, Xuehui and Li, Ying and Zhang, Zhaoyang},
  year = 2024,
  month = jul,
  journal = {IEEE Wireless Commun. Lett.},
  volume = {13},
  number = {7},
  pages = {1963--1967},
  issn = {2162-2345},
  doi = {10.1109/LWC.2024.3397668},
  urldate = {2026-02-28},
  langid = {english}
}

@article{liuInterleavedBlockSparseTransform2025,
  title = {Interleaved {{block-sparse transform}}},
  author = {Liu, Lei and Wang, Ming and Li, Shufeng and Chi, Yuhao and Wei, Ning and Zhang, Zhaoyang},
  year = 2025,
  month = apr,
  journal = {IEEE Commun. Lett.},
  volume = {29},
  number = {4},
  pages = {739--743},
  issn = {1558-2558},
  doi = {10.1109/LCOMM.2025.3542388},
  urldate = {2026-04-06},
  langid = {english}
}

@article{liuMemoryAMP2022,
  title = {Memory {{AMP}}},
  author = {Liu, Lei and Huang, Shunqi and Kurkoski, Brian M.},
  year = 2022,
  month = dec,
  journal = {IEEE Trans.  Inf. Theory},
  volume = {68},
  number = {12},
  pages = {8015--8039},
  issn = {1557-9654},
  doi = {10.1109/TIT.2022.3186166},
  urldate = {2026-04-07}
}

@article{liuRandomMultiplexing2026,
  title = {Random {{multiplexing}}},
  author = {Liu, Lei and Chi, Yuhao and Huang, Shunqi and Zhang, Zhaoyang},
  year = 2026,
  journal = {IEEE Trans.  Inf. Theory},
  pages = {1--1},
  issn = {1557-9654},
  doi = {10.1109/TIT.2026.3653055},
  urldate = {2026-02-28},
  langid = {english}
}

@article{maOrthogonalAMP2017b,
  title = {Orthogonal {{AMP}}},
  author = {Ma, Junjie and Ping, Li},
  year = 2017,
  journal = {IEEE Access},
  volume = {5},
  pages = {2020--2033},
  issn = {2169-3536},
  doi = {10.1109/ACCESS.2017.2653119},
  urldate = {2026-04-07}
}

@article{ranganVectorApproximateMessage2019a,
  title = {Vector {{approximate message passing}}},
  author = {Rangan, Sundeep and Schniter, Philip and Fletcher, Alyson K.},
  year = 2019,
  month = oct,
  journal = {IEEE Trans.  Inf. Theory},
  volume = {65},
  number = {10},
  pages = {6664--6684},
  issn = {1557-9654},
  doi = {10.1109/TIT.2019.2916359},
  urldate = {2026-04-07}
}

@ARTICLE{yang2026complexity,
  author={Yang, Jie and Hu, Wanchen and Jiang, Yi and Li, Shuangyang and Wang, Xin and Ng, Derrick Wing Kwan and Caire, Giuseppe},
  journal={IEEE J. Sel. Areas Commun.}, 
  title={{Complexity}-scalable Near-Optimal Transceiver Design for Massive {MIMO-BICM} Systems}, 
  year={2026},
  volume={44},
  number={},
  pages={3557-3574},}

@article{yang2025achievable,
  title = {Achievable {{rate maximization}} in {{MIMO-BICM systems}}: {{A unified transceiver design}}},
  author = {Yang, Jie and Hu, Wanchen and Jiang, Yi and Li, Shuangyang and Wang, Xin},
  year = 2025,
  journal = {IEEE Trans. Signal Process.},
  volume = {73},
  pages = {3346--3361},
  issn = {1941-0476},
  doi = {10.1109/TSP.2025.3592835},
  urldate = {2026-04-06},
}

@inproceedings{RFTN,
  title={{Random} Faster-than-{Nyquist} Signaling},
  author={Li, Shuangyang and Çakmak, Burak and Caire, Giuseppe and Yuksel, Melda and Elisa Conti},
  booktitle={IEEE Int. Symp. Inf. Theory, accepted},
  year={2026},
}

@article{yan2026capacity,
  title={{Capacity}-Region-Achieving Sparse Regression Codes for {MIMO} Multiple-Access Channels},
  author={Yan, Hao and Liu, Lei and Liu, Yuhao and {\c{C}}akmak, Burak and Caire, Giuseppe},
  journal={arXiv preprint arXiv:2604.11062},
  year={2026}
}

@article{huParametricMIMOOFDMChannel2025a,
  title = {{Parametric MIMO-OFDM} Channel Estimation: {A} Quasi Neural Network Approach},
  author = {Hu, Wanchen and Yang, Jie and Liang, Xin and Ran, Rong and Jiang, Yi and Zhu, Yu},
  year = 2025,
  month = dec,
  journal = {IEEE Trans. Commun.},
  volume = {73},
  number = {12},
  pages = {14931--14944},
  issn = {1558-0857},
  doi = {10.1109/TCOMM.2025.3594785},
  urldate = {2026-01-24}
}

@misc{38901,
    author= {{3GPP TS 38.901}},
    title = {{Study on channel model for frequencies from 0.5 to 100 GHz}},
    note  = {3rd Generation Partnership Project; Technical Specification Group Radio Access Network, 2020},
}

@article{tanakaStatisticalmechanicsApproachLargesystem2002,
  title = {A Statistical-Mechanics Approach to Large-System Analysis of {{CDMA}} Multiuser Detectors},
  author = {Tanaka, T.},
  year = 2002,
  month = nov,
  journal = {IEEE Trans.  Inf. Theory},
  volume = {48},
  number = {11},
  pages = {2888--2910},
  issn = {1557-9654},
  doi = {10.1109/TIT.2002.804053},
  urldate = {2026-04-07}
}

@article{tulinoSupportRecoverySparsely2013,
  title = {{Support Recovery with Sparsely Sampled Free Random Matrices}},
  author = {Tulino, Antonia M. and Caire, Giuseppe and Verd{\'u}, Sergio and Shamai, Shlomo},
  year = 2013,
  month = jul,
  journal = {IEEE Trans.  Inf. Theory},
  volume = {59},
  number = {7},
  pages = {4243--4271},
  issn = {1557-9654},
  doi = {10.1109/TIT.2013.2250578},
  urldate = {2026-04-07}
}

\end{document}